\documentclass[11pt]{article}
\usepackage{hyperref}
\pdfoutput=1
\begin{document}
\title{Compressible Flows in Fluidic Oscillators}
\author{D. Hirsch, E. C. Graff, M. Gharib \\
\\\vspace{6pt} Graduate Aerospace Laboratories, \\ California Institute of Technology, Pasadena, CA 91125, USA}
\maketitle
\begin{abstract}
We present qualitative observations on the internal flow characteristics of fluidic oscillator geometries commonly referred to as sweeping jets in active flow control applications. This is part of the fluid dynamics videos.
\end{abstract}
\section*{Description}
In the video we present schlieren visualization of the internal flow of a fluidic oscillator which creates a laterally sweeping jet without any moving parts. The exit nozzle dimensions are 0.2" wide by 0.1" thick. Compressed air is the fluid inside the jet, and it blows into ambient.

The schlieren system is comprised of two 6" refractor telescopes and a blue LED light source. Images were taken with an IDT Y7 camera at 1920x1080 pixels at 9000 Hz.

To visualize the slower speeds, ``canned air'' was used as a cold gas to fill the hose leading to the fluidic oscillator. Eventually the external jet reaches supersonic speeds and the temperature difference is no longer necessary. In the video we observe startup behavior as the valve is suddenly opened through supersonic exit velocities.
\end{document}